\documentclass[%
prd
 ,twocolumn%
 ,secnumarabic%
,amssymb, amsmath,nobibnotes,showpacs]{revtex4}
\usepackage[]{graphicx}
\usepackage{bm}%
\expandafter\ifx\csname package@font\endcsname\relax\else
 \expandafter\expandafter
 \expandafter\usepackage
 \expandafter\expandafter
 \expandafter{\csname package@font\endcsname}%
\fi

\begin{document}

\title{Semi-Classical Isotropization of the Universe during a de Sitter phase}%

\author{Francesco Cianfrani$^{12}$, Giovanni Montani$^{234}$, Marco Muccino$^2$}%
\email{francesco.cianfrani@icra.it}
\affiliation{$^{1}$ICRA-International Center for Relativistic Astrophysics.\\ 
$^2$Dipartimento di Fisica,Universit\`a  di Roma ``Sapienza'', Piazzale Aldo Moro 5, 00185 Roma, Italy.\\ 
$^{3}$ENEA Centro Ricerche Frascati (Unit\`a Fusione Magnetica), Via Enrico Fermi 45, 00044 Frascati, Roma, Italy.\\
$^{4}$ICRANet Coordinating Center Pescara, Piazzale della Repubblica, 10, 65100 Pescara, Italy.}
\date{July 2010}%

\begin{abstract}
Semi-classical states for the Wheeler-DeWitt equation of a Bianchi type I model in the presence of a scalar field are analyzed. It is outlined how this scheme can effectively describe more general situations, where the curvature of the Bianchi type IX model and a proper potential term for the scalar field are present. The introduction of a cosmological constant term accounts for the quasi-isotropization mechanism which bridges the proposed framework with a late isotropic phase. This result makes the semi-classical Bianchi I model a plausible scenario for the Universe pre-inflationary phase.  
\end{abstract}


\maketitle

\section{Introduction}

The Standard Cosmological Model (SCM) relays on the highly symmetric Friedman-Robertson-Walker (FRW) 
line element and on the existence of an inflationary phase,
taking place expectantly at $T<\mathcal{O}(10^{14}-10^{15}GeV)$.
Before this time, the thermal history of the Universe enters
a rather obscure stage, in which the thermal equilibrium
is not clearly settled down and quantum corrections
start to be significant. It is a reliable conjecture
to infer that this phase of the Universe evolution be
characterized by more general space-time structures than the
isotropic FRW model, up to suggesting that the generic
cosmological solution must be taken as the natural
arena for primordial physics.

A generic inhomogeneous cosmology admits no
isometry group and it is therefore the most suitable
scenario for a predictive quantum cosmological model.
In fact, the quantum fluctuations of the matter and geometry
fields have to preserve the causality prescription and
therefore are expectantly incompatible with a restrictive
global space-time symmetry.

As shown in \cite{BKL82}, the asymptotic evolution of the Universe toward
the initial singularity, is characterized by a point-like time evolution,
closely resembling the one of the oscillatory regime
of the Bianchi type VIII and IX models, the so-called
Mixmaster Universe \cite{BKL70,M69}.
Such a dynamical regime exhibits chaotic
properties in the vacuum case, but if we
define an average scale factor as the geometrical 
average $R(t)\sim (abc)^{1/3}$, $a,b,c$ being the
scale factors along three independent directions,
then the local character of the oscillatory
regime can take a precise physical meaning.
We can define the horizon size $d_H$ associated to
$R\sim t^{1/3}$, $t$ being the synchronous time
(we recall that the piecewise Kasner-like evolution
of the Mixmaster implies $abc\sim t$)
as $d_H\sim t$. Furthermore if $l$ is the typical
length-scale of the space metric, we can introduce
the physical averaged inhomogeneous scale as
$l_{ph} = R(t)l$. Hence the possibility to
locally recover the Mixmaster behavior on
the horizon size is associated to the request that
the physical inhomogeneous scale be much larger
than the causal regions, {\it i.e.}
$l_{ph}\gg d_H$, or equivalently $l\gg t^{2/3}$.
This picture is commonly named as the inhomogeneous
Mixmaster dynamics \cite{BKL82,K93,M95} and it
can be extended toward the quantum picture \cite{K92,IM03},
by inferring a long wavelength approximation,
which allows to neglect the role of the spatial
gradients in the dynamics, so reducing the Wheeler
superspace to the product of point-like minisuperspaces.
Thus the primordial Universe is properly
represented by Bianchi type IX model. 

In this work we investigate the possibility that the pre-inflationary phase of the Universe can be modeled by a semi-classical Bianchi type I model in the presence of a scalar field. In particular, our approach is based on the construction of the appropriate semi-classical states and on the analysis of their evolution according with the Wheeler-DeWitt (WDW) equation \cite{WdW} for the considered cosmological model. The classical Kasner-like dynamics is reproduced via the behavior of expectation values. The insertion of this scenario in a generic Bianchi type IX space-time with a self-interacting scalar field is discussed. Finally, as the main issue we evaluate the quantum effects of the vacuum energy showing that a primordial cosmological constant provides the isotropization of the model, thus connecting a Bianchi type IX model with a highly isotropic scheme.

\section{Basic statements}

When describing the early Universe dynamics, it is convenient to distinguish between the variables $\alpha$, the isotropic component, and $\beta_{ab}$, the anisotropies. In the ADM decomposition, the metric of a generic cosmological model \cite{rev} can be written in the form
\begin{equation} 
\label{metromogenea} ds^2 = N^2(t)dt^2 - e^{2\alpha}(e^{2\beta})_{ab}\,\omega^a \otimes \omega^b\ ,
\end{equation}

where $\alpha$, $N$ and $\beta_{ab}$ are space-time functions, while $\omega^a$ ($a=1,2,3$) denote the 1-forms of the spatial metric. The matrix $\beta_{ab}$ is taken diagonal and with a vanishing trace, and it has only two independent components, the so-called \emph{Misner variables} $\beta_{\pm}$, which are defined in terms of $\beta_{ab}$ as follows
\begin{eqnarray} 
\nonumber \beta_{11} &=& \beta_+ + \sqrt{3}\beta_-\\
\label{misner} \beta_{22} &=& \beta_+ - \sqrt{3}\beta_-\\
\nonumber \beta_{33} &=& - 2\beta_+. 
\end{eqnarray}

When all spatial gradients are neglected, the super-Hamiltonian constraint in the presence of a scalar field is given by
\begin{equation}
-p_{\alpha}^2 + p_+^2 + p_-^2 + p_{\phi }^2
+ e^{4\alpha}U(\beta _+,\beta _-) + e^{6\alpha} V(\phi ) = 0
\, ,
\label{inHam}
\end{equation}

in which, $p_{\alpha}$ and $p_\pm$ are the conjugate momenta to $\alpha$ and $\beta_\pm$\,, respectively, while $V$ denotes the interaction potential of the scalar field $\phi$. 

The presence of the potential $U(\beta _+,\beta _-)$
is due to the spatial curvature of the specific model
and it is negligible when the Kasner-like regime holds (Bianchi I model), while in the Bianchi IX case it reads as
\begin{eqnarray}
U(\beta_+\beta_-)=e^{-8\beta_+} + e^{4(\beta_+ + \sqrt{3} \beta_-)} + e^{4(\beta_+ - \sqrt{3} \beta_-)} -\nonumber\\
- 2 \left[ e^{4\beta_+} + e^{-2(\beta_+ + \sqrt{3} \beta_-)} + e^{-2(\beta_+ - \sqrt{3} \beta_-)} \right]\ . 
\end{eqnarray}

It is a well-known fact \cite{BK73} that the
presence of free scalar field in the Mixmaster scenario
destroy the chaotic evolution of the Universe toward the
singularity, in favor of a finite sequence of
oscillations, ending in a stable Kasner regime.

But for the real primordial Universe this is just the case.
In fact, during a Kasner regime, the momentum
conjugate to $\phi$ is constant,
say $p_{\phi} = p = const$, while
$\phi$ behaves linearly in the isotropic variable
$\alpha$, {\it i.e.} $\phi = q\alpha$, $q = const. < 1$.
Hence the condition to neglect asymptotically the interaction potential reads as
\begin{equation}
\lim _{\alpha \rightarrow -\infty}
e^{6\alpha} V(q\alpha ) = 0
\, ,
\label{cond}
\end{equation}

where we recall that the relation between the
variable $\alpha$ and the synchronous time stands
as $\alpha = (1/3)\ln t$. Thus, for a potential profile
less stiff than $V\sim e^{-6\phi /q}$, the free scalar
field approximation is well fulfilled.
The most common inflationary potentials \cite{revinfl}
are in agreement with the requirement above and
we can thus reliably infer that the singularity
is reached by a finite sequence of Kasner regimes.

However, a more subtle question concerns the
possibility that the vacuum energy associated to
the inflationary phase transition (and corresponding
to $V\sim \rho _{\Lambda}\sim const.$) can play
a role already during the oscillatory regime.
Such a picture would correspond to the request
that the vacuum energy be few orders of magnitude greater than
the Grand Unification scale ($\sim 10^{14}\div10^{15} GeV$). If we introduce a
typical scale via the relation
$l = 1/\sqrt{\chi \rho _{\Lambda}}$,
$\chi$ being the Einstein constant, the
conditions to deal with a primordial inflation,
starting during the oscillatory regime, are summarized
by the following analysis.
i)-The de Sitter phase starts when
$1/t^2\sim \chi \rho _{\Lambda}$, {\it i.e.} the vacuum
energy must be source of the dominant geometrical
contribution, coming from the time derivatives of
the 3-metric. By other words, inflation begins when
$l \sim d_H\sim t$.
ii)-The slow-rolling evolution of the scalar
field \cite{KT90}, which is well-represented by
the emergence of a vacuum energy contribution,
can proceed only if the spatial gradients are
sufficiently small, namely we must have
$l \ll l_{ph}$. The de Sitter phase lives
at the horizon scale where the Universe is essentially
smooth.
iii)-The idea that inflation starts during a
Kasner phase of the oscillatory regime requires that
that the vacuum term dominates the Mixmaster
potential, so preventing the transition to a
later Kasner period. Thus, to deal with an
exponential expansion, we need that
the following restriction be fulfilled
$\rho _{\Lambda}\gg e^{-2\alpha }\bar{U}$,
$\bar{U}$ being the average value of the potential $U$
over the domain cut by the Kasnerian constant of
motion $p_+^2 + p_-^2$, {\it i.e.} over the region
$e^{4\alpha}U(\beta _+,\beta_-) \sim p_+^2 + p_-^2$.
This condition eventually stands
$l \ll \bar{l}_{ph} = \bar{l}t^{1/3}$,
$\bar{l} = 1/\sqrt{\chi \bar{U}}$ being a typical
length characterizing the Mixmaster potential.
Because $U\propto 1/l^2$, 
this length scale $\bar{l}$ is proportional to
$l$ itself by a factor greater than the unity.
For the inhomogeneous case,
this last condition is redundant with respect to that one
fixed at the point ii), but, 
in the homogeneous sector treated below,
it holds alone.

Thus, at the end of this analysis we can claim that
inflation starts during the oscillatory regime
only if $l \sim d_H\ll l_{ph}$, which is nothing
more than the existence condition for the inhomogeneous
Mixmaster. In the homogeneous case, this same
request takes the form 
$l \sim d_H\ll \bar{l}_{ph}$, which corresponds
to the condition to deal with a Kasner regime.

We can conclude that the only requirement to deal with
a very primordial inflation, emerging from the
inhomogeneous Mixmaster, is that
a transition phase takes place at sufficiently high temperature
of the Universe, say $l \sim x l_{Pl}$,
where $l_{Pl}\sim 10^{-33}cm$ is the Planck length
and $x\in (1, 10^{3})$. The current limit coming from the isotropic scheme
is $x>\mathcal{O}(10^{3})$ (see \cite{KDM} and references therein).

However in \cite{KM97} it has been show that a classical
limit for the gravitational field dynamics can not
take place before the Mixmaster ends and therefore
the scenario we are addressing here requires that inflation
emerges from a quantum (or a semiclassical) phase
of the Universe. Thus, despite its apparent simplicity,
the following Hamiltonian constraint
\begin{equation}
-p_{\alpha}^2 + p_+^2 + p_-^2 + p_{\phi }^2
+ e^{6\alpha} \rho _{\Lambda} (x^{\gamma}) = 0
\,
\label{inHam1}
\end{equation}

properly describes a real phase of the early Universe
evolution. In the present analysis we always
removed the thermal bath of ultrarelativistic particles,
having an energy density of the form
$\rho _r = f^2(x^{\gamma}) e^{-4\alpha}$, $f$ being a generic space function.
This term does not alter the oscillatory phase
\cite{BKL82} because it is a test term during the
Kasner regime \cite{LL}. The possibility to
neglect the ultrarelativistic energy with respect to
the vacuum one is natural in the SCM, since the
transition phase is mediated via the coupling of
the scalar field to the thermal bath, {\it i.e.}
the effective potential is that one calculated
at finite temperature. This same picture
is expectantly till valid in the generic inhomogeneous
case, at least within each horizon.

\section{Wheeler-DeWitt quantization}

The canonical quantum dynamics is implemented
by the requirement that the constraint (\ref{inHam})
is translated into an operator annihilating the
local state function $\psi _x(\alpha, \beta _{\pm})$.
The Universe wavefunctional is then obtained as
the infinite product of state functions taken on
independent horizons, say
$\Psi = \Pi_x \psi _x$. To avoid many of
the subtle questions concerning the inhomogeneous
functional sector, we address our main goal,
the possibility that the vacuum energy isotropizes
the Universe on a quantum or a semiclassical level
(the classical case has been successfully treated
in \cite{KM02}), by the analysis of the homogeneous Bianchi IX
cosmological model.
Apart from the heuristic character of the long-wavelength
approximation and some technicalities concerning
the supermomentum constraint, we are really
confident that the simplified homogeneous
analysis already contains all the physical
ingredients to qualitatively describe
the sub-horizon physics even in the generic
inhomogeneous case.

Let us now investigate the quantum dynamics of the Bianchi IX model, starting from the Bianchi type I.

At first, momenta are replaced by derivatives in the conjugate coordinates. Then, the super-Hamiltonian constraint $H=0$ leads to the Wheeler-DeWitt equation (WDW), {\it i.e.}
\begin{equation}  
\label{WDWI} \left[ e^{c\alpha} \partial_{\alpha}\!\left(\!e^{-2b\alpha} \partial_{\alpha} e^{c\alpha}\! \right)\! - e^{-3\alpha}\!\Delta + e^{3\alpha} V(\!\phi\!)\!\right] \Psi_I(\!\alpha,\beta^r)\!=\! 0
\end{equation}

in which the Laplacian in the variables $\beta^r\!=\!\beta_{\pm},\phi$ is
$$ \Delta = \partial^2_+ + \partial^2_- + \partial^2_{\phi}\ . $$

In Eq. (\ref{WDWI}) we wrote a symmetric super-Hamiltonian operator, introducing generic parameters \emph{b} and $c\!=\!b\!-\!\frac{3}{2}$\,.
Such a choice for the operator ordering is not the most general one, but it captures several cases. Moreover, we will emphasize that the value of the parameter $b$ does not affect the proposed semi-classical picture.

Approaching the singularity the scalar field potential term can be neglected
and the solution is given by 
\begin{eqnarray} 
\label{kg3} \Psi_I(\alpha,\beta_\pm,\phi) =e^{\frac{3}{2}\alpha} \int dk_+dk_-dk_\phi \sqrt{\frac{2}{3\,K_k}} \nonumber\\(a_k\,e^{\frac{3}{2}\imath K \alpha +\imath k_r\beta^r}
+b_k\,e^{-\frac{3}{2}\imath K_k \alpha + \imath k_r\beta^r }), 
\end{eqnarray}
in which $k_r\!=\{k_\pm,k_\phi\}$, while $a_k$ and $b_k$ denote weights of the Fourier expansion. The conjugate momentum to the variable $\alpha$\, is
\begin{equation}\label{kappa}
K=\frac{2}{3}\sqrt{\varepsilon^2 - b^2}\ \ \ \ ,\ \ \ \ \displaystyle \varepsilon^2=k^2_+ +k^2_-+k_\phi^2\ . 
\end{equation}

It is worth noting that the parameter $b$ labeling the operator ordering enters merely the definition of $K$. This fact means that $b$ fixes the interval in the $\epsilon$-line where the solution has an oscillatory behavior. Because, we are interested in developing wave-packets, we assume wave-functions to be negligible for $\epsilon\lesssim b$. In this regime, expectation values have actually no $b$-dependence.   

The Hilbert space is $\mathcal{L}^2(\beta^r,d\mu)$, where the scalar product is given by
\begin{equation} <\psi_2|\psi_1>=\frac{\imath}{2}\int e^{-3\alpha}\left[ \Psi_2^{*} (\partial_{\alpha} \Psi_1) - (\partial_{\alpha} \Psi_2^{*})\,\Psi_1 \right] d\beta_+d\beta_-d\phi   \end{equation}

which is positive-defined as far as the proper separation of frequencies occurs by fixing $a_k\!=\!0$\, (such that the case of an expanding Universe is selected out). 

\section{Semi-classical states}

The localization of the solution (\ref{kg3}) in the phase space at a given initial position $\beta^r_0$ is achieved by the following Gaussian wave packets
\begin{equation} \label{pacchetto phi} b_k = \frac{1}{(2\pi\,\sigma^2)^{\frac{3}{4}}} \exp \left( -\frac{\delta k_+^2 + \delta k_-^2 + \delta k_\phi^2}{4\,\sigma^2} \right) e^{-ik_r\beta^r_0}\,,\end{equation}    
where  $\delta k_r = k_r - \bar{k}_r$, while the variances have the same values $\sigma$\,.

The evolution of wave-packets is investigated by virtue of a saddle point expansion around $\bar{k}_r$ and a proper semi-classical behavior is found.  In particular, the evaluation of expectation values and variances gives 
\begin{eqnarray} 
\label{aspett1phi} \langle \beta^r \rangle = \beta^r_0- \bar{v}_r\,\alpha + \mathcal{O}(\bar{\varepsilon}^{-\frac{3}{2}}),\qquad \Sigma_r \bar{v}_r^2=1,\\ 
\label{aspett2phi} \langle{\Delta\beta^r}^2 \rangle = \frac{1}{4\sigma^2} + \frac{\sigma^2\,\alpha^2}{\bar{\varepsilon}^2} + \mathcal{O}(\bar{\varepsilon}^{-3}).
\end{eqnarray}

Hence, for sufficient high values of $\bar{\varepsilon}=\varepsilon(\bar{k}^r)$ the behavior of expectation values is approximated by the Kasner-like dynamics. The restriction to high values of $\bar{\varepsilon}$ corresponds to the well-known result that the semi-classical picture is inferred only for high values of the initial momenta. In what follows, we will consider the corrections of the $\bar{\varepsilon}^{-\frac{3}{2}}$ order to be negligible. 

In this scheme a measure of the spread of wave packets is given by the ratio of the square root of variances with $\langle \beta^r \rangle$. In particular, such a quantity at late times goes as
\begin{equation}
\frac{\sqrt{\langle{\Delta \beta^r}^2 \rangle }}{\langle \beta^r \rangle} \approx \frac{\sigma}{\bar{v}_r\,\bar{\varepsilon}} + \mathcal{O}(\bar{\varepsilon}^{-\frac{3}{2}}) \ ,
\end{equation}

such that the wave-function spread tends to a constant value. This fact implies that proper initial conditions can be chosen for gaussian wave-packets such that they remain well-localized around expectation values and the semi-classical picture holds. 

Therefore, Gaussian wave functions are proper semi-classical states for the Bianchi type I model in the presence of the scalar field.

\subsection{Bianchi IX model}

The WDW equation for a Bianchi type IX model is modified by the presence of the 3-dimensional scalar curvature, which acts as a potential term and whose expectation value on semi-classical states gives 
\begin{eqnarray}
\nonumber \lim_{\alpha\rightarrow -\infty}\langle e^{4\alpha}U \rangle \propto\lim_{\alpha\rightarrow -\infty} e^{\frac{8}{\sigma^2}} [ e^{4\alpha(1+2\bar{v}_+)} + \\
\label{nullo} + e^{4\alpha(1-\bar{v}_+ - \sqrt{3}\bar{v}_-)} +  e^{4\alpha(1-\bar{v}_+ + \sqrt{3}\bar{v}_-)} ] + \mathcal{O}(\bar{\varepsilon}^{-\frac{3}{2}}) \ .\end{eqnarray}
This expression approaches $0$ when
\begin{equation} \label{fifff} \left\{ \begin{array}{lll}
1 + 2\,\bar{v}_+ > 0 \vspace{0.1cm}\\
1 - \bar{v}_+ - \sqrt{3}\,\bar{v}_- > 0 \vspace{0.1cm}\\
1 - \bar{v}_+ + \sqrt{3}\,\bar{v}_- > 0
\end{array} \right. \ \ \ \ \Rightarrow\ \ \ \ 
\left\{ \begin{array}{lll}
\bar{v}_+^2 < \frac{1}{4} \vspace{0.1cm}\\
\bar{v}_-^2 < \frac{1}{12} \vspace{0.1cm}\\
\frac{2}{3} <\bar{v}_{\phi}^2 < 1
\end{array} \right. \ .\end{equation}

The conditions above coincide with the relations found in a classical framework to remove the chaotic behavior \cite{B00}. Hence, to restrict the domain of the parameters $\bar{v}_\pm$, $\bar{v}_\phi$ according with inequalities (\ref{fifff}) guarantees that both the classical and the semi-classical dynamics of the Bianchi type IX model resembles that of a Bianchi type I space. 

Therefore, the obtained results support the idea that the proposed scenario realizes a proper semi-classical description of the Early phases of the Universe.  

\section{Quasi-isotropization mechanism}

A quasi-isotropization mechanism is required in order to suppress anisotropies, so reconciling the early Universe dynamics with its late evolution. Inflation can provide such a suppression \cite{KM02} on a classical level. Here we are going to realize the inflationary phase via the introduction of a scalar field (the inflaton), which acquires a non-vanishing vacuum expectation value modeled by a cosmological constant $\rho_\Lambda$.
 
The WDW equation associated with a Bianchi type I model in the presence of a scalar field and of a cosmological constant is given by (we fix $c=0$, because as in the previous case a different operator ordering does not provide any significant modification to the quasi-isotropization mechanism)
\begin{equation} \label{9lambda} e^{-3\alpha} \left[ \partial_{\alpha}^2 - 3 \partial_{\alpha} - \Delta + e^{6\alpha}\rho_\Lambda \right] \Psi(\alpha,\beta_{\pm},\phi) = 0 \ .\end{equation}
The solution of such an equation restricted to negative frequencies has the following form
\begin{equation} 
\label{soluzionea} \Psi(\alpha,\beta^r)\!=\int dk_+dk_-dk_\phi b_k \frac{\Gamma(1\!+\!n_k)}{\sqrt{N_k}} J_{n_k}[z(\rho_\Lambda,\alpha)] e^{\frac{3}{2}\alpha- \imath k_r \beta^r },
\end{equation}
in which $N_k$ is the normalization factor, $K$ retains the form (\ref{kappa}), while $\Gamma(1+n_k)$ and $J_{n_k}(z)$ denote the Gamma function and the Bessel function of the first kind, respectively, where $n_k\!=\!-\frac{\imath}{2}K$ and $z(\rho_\Lambda,\alpha)\!=\!\frac{\sqrt{\rho_\Lambda}}{3} e^{3\alpha}$. 

Eq. (\ref{9lambda}) aims to describe the phase of the Universe when the transition from the anisotropic to the isotropic regime takes place. In order to characterize such a transition, the two relevant cases $z\ll1$ and $z\gg1$ are going to be discussed separately.  

\subsection{Early phase}

Let us consider the early phase, where $ 
z=\frac{\sqrt{\rho_\Lambda}}{3}\,e^{3\alpha}\ll1$
and the Bessel function can be expanded as follows \cite{AS}
\begin{equation} \label{bessel} J_n(z)=\sum^{+\infty}_{l=0} \frac{\left( \frac{z}{2} \right)^{2l+n}}{\Gamma(1+l+n)\,l!}\ \longrightarrow \frac{\left( \frac{z}{2} \right)^{n}}{\Gamma(1+n)}\ ,\end{equation}
such that the solution (\ref{soluzionea}) can be approximated by the following asymptotic form
$$ \Psi(\alpha',\beta^r) =e^{\frac{3}{2}\alpha'} \int dk_+dk_-dk_\phi \sqrt{\frac{2}{3\,K}} b_k\,e^{- \imath \left( \frac{3}{2} K_k \alpha' + k_r\,\beta^r \right)}. $$

The expression above coincides to the solution of the Bianchi type I case in the presence of a scalar field (\ref{kg3}) in terms of the re-defined isotropic variable $\alpha'$, which is given by
\begin{equation} 
\label{smalla} \alpha' = \alpha + \frac{1}{3}\ln \frac{\sqrt{\rho_\Lambda}}{6}.
\end{equation}
The wave packets can be developed according with the procedure adopted in the previous cases 
and all quantities are now functions of $\alpha'$.

The evaluation of expectation values for operators corresponding to phase-space coordinates can be carried on just like in the case of a Bianchi type I model. Hence, wave-packets remain well localized around the classical trajectory.

Therefore, for $\alpha \ll \frac{1}{3} \ln \frac{3}{\sqrt{\rho_\Lambda}}$\,, the presence of the cosmological constant term $\rho_\Lambda$ does not modify significantly the semi-classical picture inferred for the early Universe dynamics, which can be described by the Bianchi type I model in the presence of a scalar field.

\subsection{Late phase}

For $z\!\gg\!1$, the Bessel functions can be approximated with the following expression \cite{AS}
$$ J_n(z) \approx \frac{1}{\sqrt{2\pi z}}\left[ e^{\imath \left( z - \frac{n\,\pi}{2} - \frac{\pi}{4} \right)} + e^{-\imath \left( z - \frac{n\,\pi}{2} - \frac{\pi}{4} \right)} \right] \ .$$
The solution of the WDW equation within this scheme turns out to be given by
\begin{equation}
\label{soluzioneb} \Psi(\alpha,\beta^r)\! =\int dk_+dk_-dk_\phi  \frac{b_k e^{-\imath(k_r\,\beta^r)}\,e^{-\imath (\frac{\sqrt{\rho_\Lambda}}{3} e^{3\alpha} - \frac{\pi}{4}- \frac{n_k}{2})}}{\sqrt{\sqrt{\rho_\Lambda}\,\sinh( \imath n_k )}}.
\end{equation}

The explicit computation of Gaussian wave-packets is performed by a saddle-point expansion around the expectation value. Finally, the behavior of the expectation values and of the variances at late time is given by
\begin{eqnarray}
\label{aspett1ll} \langle \beta^r \rangle  = \beta^r_0+\mathcal{O}(\bar{\varepsilon}^{-\frac{3}{2}}),\qquad \langle{\Delta \beta^r}^2\rangle  = \frac{1}{4\sigma^2} + \mathcal{O}(\bar{\varepsilon}^{-3}).
\end{eqnarray}

It is worth noting that in the adopted approximation scheme the evaluation of $\sqrt{\langle{\Delta \beta^r}^2 \rangle }/\langle \beta^r \rangle$ gives 
\begin{equation}
\frac{\sqrt{\langle{\Delta \beta^r}^2 \rangle }}{\langle \beta^r \rangle} \approx \frac{1}{4\sigma^2\beta^r_0} + \mathcal{O}(\bar{\varepsilon}^{-\frac{3}{2}}) .
\end{equation}

The quantity above does not depend on the time-like variable and it can be set smaller than any given quantity by a proper choice of initial conditions $\beta^r_0$ and $\sigma$. This fact implies that the semi-classical picture is consistent with the adopted approximation scheme. 

Furthermore, the expectation values of anisotropies freeze out to constant values, corresponding to the chosen initial conditions. Because constant $\beta$ can always be avoided by a re-definition of the 1-form $\omega^a$ of the spatial metric, the final space does not contain anisotropies and this scenario offers a bridge between a Bianchi type I model and a late isotropic phase.

Hence, the quasi-isotropization mechanism works also in the semi-classical regime. Therefore, a cosmological constant term can determine the isotropization of the Universe, even if the spontaneous symmetry breaking process (by which a vacuum energy expectation value arises) takes place during the quantum phase of the Universe.

\section{Concluding Remarks} 

In this work we focused on the semi-classical behavior of a homogeneous Bianchi type I model in the presence of a scalar field and of a cosmological constant term, which mimic an inflationary scenario. At first, we discussed the relevance that the proposed model had in the early phase of the Universe evolution and we pointed out how the Bianchi type I model captured the main features of the general cosmological solution. 
The semi-classical analysis of the WDW equation was performed via the definition of gaussian wave packets, and the consistency of such a scheme was confirmed by the analysis of the spread around expectation values for the operators associated with the anisotropies and the scalar field. 

We did not enter into the details of the spontaneous symmetry breaking mechanism responsible for the transition to an accelerated expansion behavior and we modeled the inflationary phase by the potential of the inflaton scalar field, taken as a primordial vacuum energy $\rho_\Lambda$. Hence, we expect that the proposed picture describes qualitatively all type of the inflationary scenarios realized via a scalar field. From the investigation of the expectation values of the configuration variables, we concluded that: i)the semi-classical approximation worked during the accelerated expansion phase because wave-packets remained well-localized around their expectation values; ii)the expectation values of anisotropic variables were suppressed by the cosmological constant term to negligible values at late times. Therefore, we can conclude that the inflationary phase could bridge a primordial Bianchi type I dynamics with a late isotropic phase. This result can be regarded as appealing, because it confirms that no matter inflation occurs after or before the quantum to classical transition, nevertheless it provides the quasi-isotropization mechanism needed to reconcile the early and late phases of the Universe. 

Having demonstrated the plausibility of a semi-classical inflationary phase, the next step involves the study of the evolution for scalar and tensor perturbations, whose quantum character is here obvious. Although this analysis is expected to depend on the inflationary paradigm adopted, nevertheless it would allow us to compare such a picture with Cosmic-Microwave-Background data and to infer a quantitative characterization for the model.

\end{document}